\documentclass[11pt]{article}

\usepackage{psfrag}
\usepackage{graphicx}

\newcommand{\eq}[1]{equation~(\ref{#1})}
\newcommand{\eqs}[2]{equations~(\ref{#1}) and~(\ref{#2})}

\newcommand{\fig}[1]{fig.~(\ref{#1})}
\newcommand{\figs}[2]{figs~(\ref{#1}) and~(\ref{#2})}

\newcommand{\mpl}{m_{\mbox{\tiny pl}}}
\newcommand{\phiinst}{\phi_{\mbox{\tiny inst}}}
\newcommand{\phiend}{\phi_{\mbox{\tiny end}}}
\newcommand{\Einst}{E_{\mbox{\tiny inst}}}

\newcommand{\lsim}{\mathrel{\rlap{\lower 4pt
    \hbox{\hskip 1pt $\sim$}}\raise 1pt \hbox {$<$}}}
\newcommand{\gsim}{\mathrel{\rlap{\lower 4pt
    \hbox{\hskip 1pt $\sim$}}\raise 1pt \hbox {$>$}}}

\begin{document}
\thispagestyle{empty}

\mbox{}

{\raggedleft
\footnotesize
gr-qc/9711035 \\
BROWN-HET-1099\\
WU-AP/67/97\\}

\vspace{3mm}

\begin{center} 	{\Large\bf\expandafter{
      Chaotic Dynamics and Two-Field Inflation\\}}
\end{center}

\vspace{5mm}

\begin{center}
{\Large Richard Easther}
\footnote{easther@het.brown.edu}
\medskip

Department of Physics,  Brown University, \\
Providence, RI  02912, United States of America.
\bigskip

{\Large Kei-ichi Maeda}
\footnote{maeda@mn.waseda.ac.jp}

\medskip
Department of Physics,  Waseda University, \\
 3-4-1  Okubo, Shinjuku-ku, Tokyo, Japan.

\bigskip

\end{center}

\vspace{5mm}

\setcounter{footnote}{0}
\setcounter{page}{0}

\section*{Abstract}

We demonstrate the existence of chaos in realistic models of two-field
inflation.  The chaotic motion takes place after the end of inflation,
when the fields are free to oscillate and their motion is only lightly
damped by the expansion of the universe. We then investigate whether
the presence of chaos affects the predictions of two-field models, and
show that chaos enhances the production of topological defects and
renders the growth rate of the universe sensitively dependent upon the
``initial'' conditions at the beginning of the oscillatory era.
\vfill

\newpage

\section{Introduction}

Inflationary cosmology and chaos theory are sites of intense activity
within astrophysics and mathematics. As the equations governing the
evolution of an inflationary universe are nonlinear, it is natural to
ask whether the insight gained from studies of chaos in systems of
differential equations sheds any light upon the behavior of the
inflationary universe. In this paper we present {\em prima facie\/}
evidence that the dynamics of realistic inflationary cosmological
models based on two coupled scalar fields can be
chaotic.\footnote{Inflationary models with chaotic dynamics, which are
sensitively dependent upon their initial conditions, are not
necessarily synonymous with chaotic inflationary models
\cite{Linde1983b}, which are recommended, in part, by their relative
insensitivity to initial conditions.}

Chaotic dynamics in models consisting of a single (non-minimally
coupled) scalar field in a Robertson-Walker background have been
studied by several authors
\cite{CalzettaET1993a,CalzettaET1995a,BombelliET1997a,HelmiET1997a}.
Moreover, Cornish and Shellard~\cite{CornishET1997c} show that the
familiar inflationary model of a minimally coupled field with a
quadratic potential exhibits chaotic dynamics when spacetime has
positive spatial curvature.

In chaotic models with just one scalar field the scale factor must
always be an intrinsic part of the chaotic system; thus, the timescale
over which chaos becomes manifest is roughly defined by the lifetime
of the universe.  Cornish and Levin~\cite{CornishET1995a} point out
that in two-field models the oscillations of the fields can be the
primary source of chaos, so chaotic effects can become significant on
comparatively short timescales.  However, their analysis hinges on the
dependence of the final state of the universe on the initial
conditions, so their work does not fully realize the possibility that
the time-scale associated with chaos in two-field inflation can be
much less than the lifetime of the universe. Hence, while chaos has
been found in several homogeneous inflationary models, it would
typically only be detectable by an ``observer'' able to study either a
large ensemble of universes, or the same universe during more than one
cycle of expansion and contraction. A different situation arises in
analyses of reheating, when the inflaton decays into a second scalar
field \cite{KofmanET1997a,GreeneET1997a,GreeneET1997b}. At least one
of the fields is expanded in terms of its spatially dependent modes,
so the resulting dynamical system is more complex than in the
homogeneous case. However, small changes in the model parameters can
cause massive variations in the occupation numbers associated with
the modes, which is the qualitative behavior associated with chaos.

The chaotic motion we examine is derived from the homogeneous equations
of motion and takes place in a well-defined era after the end of
inflation. Consequently, the chaotic period we identify can, in
principle, have effects which are observable from within the
present-day universe. Furthermore, previously studied examples of
chaos in homogeneous inflationary models typically depend upon
non-zero spatial curvature, a cosmological constant, or a non-minimal
coupling between the fields and the gravitational sector. All these
terms are absent from our model, and the inflationary system we
analyze has already received considerable attention because of its
promising cosmological predictions.  The only additional assumption we
need is that the parameter values and couplings between the scalar
fields and the rest of the matter sector (which we do not include
explicitly) permit a sufficiently long and lightly damped oscillatory
period for the chaotic effects to become manifest.

We begin by analyzing the evolution of the fields in the limiting case
when the friction term arising from the expansion of the universe is
absent from the equations of motion for the fields. We demonstrate
that this system is chaotic, and that there is a minimum energy below
which chaos does not occur. When the friction term is restored, we
show that the energy changes slowly on a time scale defined by the
typical oscillation period in the post-inflationary era. The chaotic
properties of the system persist, but are now transient and cease to
be apparent once the energy drops below the critical value needed for
chaos.

If inflation is the source of the fluctuations that seed structure
formation, the ``initial conditions'' of the many post-inflationary
horizon volumes accessible to a present-day observer cannot have been
strictly identical. However, analyses of inflationary models usually
make the tacit assumption that small differences in field values and
velocities between post-inflation horizon volumes can be ignored when
computing ``zero order'' quantities such as the scale factor.  The
validity of this assumption cannot be taken for granted in the
presence of chaos. In particular, we consider whether chaotic motion
enhances the production of topological defects, and if there is a
well-defined relationship between the expansion of the universe and
the decrease in density during the chaotic era.

\section{The Model and the Equations of Motion}

The equations of motion for inflationary models minimally coupled to
Einstein gravity are well known, and for two scalar fields with a
combined potential $V(\phi,\psi)$ they take the form
\begin{eqnarray}
H^2  = \left(\frac{\dot{a}}{a}\right)^2 &=& \frac{8\pi}{3\mpl^2} \left[
   \frac{\dot{\phi}^2}{2}+ \frac{\dot{\psi}^2}{2} +
V(\phi,\psi)\right],\label{Hsqrd}
\\
\frac{\ddot{a}}{a} &=& \frac{8\pi}{3\mpl^2}\left[V(\phi,\psi) -
\dot{\phi}^2- \dot{\psi}^2\right],\label{addot}
\\
\ddot{\phi} &=& -3H\dot{\phi} -\frac{\partial V}{\partial\phi},
\label{phiddot} \\
\ddot{\psi} &=& -3H\dot{\psi} -\frac{\partial V}{\partial\psi},
\label{psiddot}
\end{eqnarray}
where $a$ is the scale factor of the Robertson-Walker spacetime
metric, $\mpl$ is the Planck mass, and dots denote differentiation with
respect to time. The specific model we consider is defined by the
potential
\begin{equation}
V(\phi,\psi) =
  \left( M^2 - \frac{ \sqrt{\lambda} }{2} \psi^2 \right)^2
  + \frac{m^2}{2} \phi^2 + \frac{\gamma}{2} \phi^2 \psi^2,
  \label{potential}
\end{equation}
where we assume that $\gamma\ge0$. The inflationary properties of this
potential have been widely studied
\cite{KofmanET1987a}-%
\nocite{KofmanET1988a,SalopekET1989a,Kofman1991b,Linde1991b,Linde1993a,%
CopelandET1994a,GarciaBellidoET1996a,GarciaBellidoET1996b,
GarciaBellidoET1996c}%
\cite{GarciaBellidoET1997a}.
Inflation is driven by the $\phi$ field: when $\phi$ is large, $\psi$
acquires a large positive effective mass, and the fields roll down the
$\psi\approx 0$ groove in the potential.  There are two distinct
inflationary modes, which Copeland {\em et al.\/}
\cite{CopelandET1994a} dub the {\em inflaton dominated\/} and
{\em vacuum dominated\/} regimes. In the former case, the total energy
density is much greater than $M^4$, while during vacuum dominated
expansion the energy density is of the order of $M^4$.  The degenerate
minima located at $\phi=0$ and $\psi = \pm M \sqrt{2}\lambda^{-1/4}$
(separated by a barrier of height $M^4$) can lead to the formation of
topological defects.

The inflationary dynamics associated with this potential are described
in \cite{CopelandET1994a}, and we summarize them here. The effective
mass of the $\psi$ field becomes zero when $\phi^2 = \phiinst^2$, where
\begin{equation}
 \phiinst^2 = \frac{2\sqrt{\lambda}M^2 }{\gamma}.
\end{equation}
When $\phi^2$ first drops below $\phiinst^2$, $\psi$ is balanced on an
unstable local maximum in the potential and becomes free to oscillate.
Conversely, during inflation $\psi \approx 0$ and the potential
reduces to the single field case,
\begin{equation}
V(\phi) \approx M^4 + \frac{m^2}{2} \phi^2.
\end{equation}
If inflation ends with $\phi^2\gg\phiinst^2$, the $M^4$ term can be
ignored during inflation, which ceases when
\begin{equation}
\phiend^2 \approx \frac{\mpl^2}{4\pi},
\end{equation}
and the slow roll conditions are no longer satisfied. If inflation
continues until $\phi^2 \sim \phiinst^2$, it ends quickly via the
instability in the $\psi$ field. A realistic model in which inflation
ends with the universe in the inflaton dominated stage must satisfy
\begin{eqnarray}
4\sqrt{\pi}\frac{M}{\mpl} &\ll& 3\times 10^{-3}, \\
\frac{\gamma}{\sqrt{\lambda}} &>& 8\pi \frac{M^2}{\mpl^2}, \\
m &=& \frac{5.5\times 10^{-6}}{\sqrt{8 \pi}} \mpl. \label{mconst}
\end{eqnarray}
The first inequality ensures that the $M^4$ contribution can be
ignored, the second guarantees that slow rolling is violated and
inflation ends when $\phi^2 > \phiinst^2$, and the last gives
COBE-normalized density perturbations.\footnote{The numerical value
is taken from \cite{CopelandET1994a}, and would be modified slightly
by using the four year COBE result. However, the difference will not
affect our work.}

\section{The Frictionless Case }

We are concerned with the epoch in which the fields are weakly damped
and free to oscillate. We begin our analysis by investigating the
completely frictionless system.  Physically, this is equivalent to
assuming that the universe is not expanding and that the energy
density is static. Mathematically, this means the equations of motion
for the fields can be derived from a Hamiltonian.  The equations of
motion simplify further if we rescale the variables:
\begin{equation}
\phi = \frac{M^2}{m}\Phi, \quad
\psi = \frac{M}{\lambda^{1/4}}\Psi, \quad
t = M^2T, \quad
\gamma = \Gamma \frac{\sqrt{\lambda}m^2}{M^2}.
\label{newvars}
\end{equation}
The potential becomes
\begin{equation}
V(\Phi,\Psi) = M^4\left[ \left( 1-\frac{\Psi^2}{2}\right)^2 +
 \frac{\Phi^2}{2} + \frac{\Gamma}{2} \Phi^2 \Psi^2 \right]
\label{Vham}
\end{equation}
and the equations of motion are
\begin{eqnarray}
\frac{d^2\Phi}{dT^2} &=& -\frac{1}{M^{4}} \frac{\partial V}{\partial \Phi} ,\\
\frac{d^2\Psi}{dT^2} &=& -\frac{1}{M^{4}} \frac{\partial V}{\partial \Psi}, \\
\frac{E}{M^4} &=& \frac{1}{2} \left(\frac{d \Phi}{dT}\right)^2 +
\frac{1}{2} \left(\frac{d \Psi}{dT}\right)^2 + \frac{V}{M^4},
\end{eqnarray}
where $E$ is conserved. These equations are independent of $M$, so we
fix $M=1$ for convenience. If $\Gamma=0$, we have a special case: the
equations separate and our two dimensional system has two constants of
the motion, ensuring that it is integrable~\cite{OttBK1}, and
therefore not chaotic. To our knowledge, Hamiltonian chaos associated
with the potential described by \eq{Vham} has not been considered
before. However, Matinyan and M\"{u}ller \cite{MatinyanET1996a}
discuss a system which lacks the $\Phi^2/2$ term in the potential but
is otherwise equivalent to our model in the absence of friction. In
general, Hamiltonian systems with terms like $\Phi^2\Psi^2$ are
non-integrable, and Steeb {\em et al.\/} summarize discussions of
chaotic behavior when the potential consists solely of a
$\Phi^2\Psi^2$ term \cite{SteebET1985a}.

\subsection{Applying the Toda-Brumer Test}

The Toda-Brumer test \cite{Toda1974a,Brumer1974a} checks for the
presence of a growing, unstable solution, which is typically a
prerequisite for chaos.\footnote{The Toda-Brumer criterion is
equivalent to the Gaussian curvature of the potential being
negative. However, even if $||V_{\Phi\Psi}''||$ is never negative
chaos can still occur, so the Toda-Brumer test does not provide a
sufficient condition for stability.}  The criterion for instability is
the existence of $\Phi$ and $\Psi$ for which
\begin{equation}
||V_{\Phi\Psi}''|| = \frac{d^2 V}{d\Phi^2} \frac{d^2 V}{d\Psi^2}
	- \left(\frac{d^2 V}{d\Phi d\Psi}\right)^2 < 0.
\label{TBdet}
\end{equation}
Inserting \eq{Vham} into \eq{TBdet} gives
\begin{equation}
||V_{\Phi\Psi}''|| = (\Gamma \Psi^2 + 1)(3 \Psi^2 -2) +
\Gamma\Phi^2 (1-3 \Gamma \Psi^2).
\label{TBtest}
\end{equation}
Setting \eq{TBtest} to zero and solving gives a parametric equation
for the borders of the region within which $||V_{\Phi\Psi}''||$ is
negative,
\begin{equation}
\Phi^2 = \frac{(3\Psi^2 -2)(\Gamma \Psi^2 +1)}{\Gamma(3 \Gamma \Psi^2 -
1) }.
\end{equation}
Inserting this value of $\Phi$ into $V(\Phi,\Psi)$ yields $U(\Psi)$,
the value of the potential on the boundary of the instability region,
\begin{equation}
U(\Psi) = \left(1- \frac{\Psi^2}{2}\right) + \frac{1}{2\Gamma}
\frac{(3\Psi^2 -2)(1+\Gamma\Psi^2)^2 }{ 3\Gamma\Psi^2 -1}.
\end{equation}
Minimizing $U(\Psi)$ subject to the constraint that $\Phi$ is positive
returns the minimum energy (as a function of $\Gamma$) needed for
instability. When $\Psi^2 = 2/3$, $\Phi = 0$ and $U = 4/9$, putting an
upper bound on the minimum energy, which must be non-zero because
$||V_{\Phi\Psi}''||$ is positive at the minima of the potential. More
precisely, solving $U'(\Psi) =0$ shows that $U(\Psi)$ has extremal
values at
\begin{equation}
\Psi^2 = \frac{4}{9} + \frac{1}{9\Gamma} \pm
  \frac{\sqrt{(2\Gamma-1)(2\Gamma-7)}}{9\Gamma}, \label{Psimin}
\end{equation}
where we have ignored two roots for which $\Psi$ is imaginary. A
little analysis shows that when $\Gamma\le9/2$, the minimum energy for
instability is $4/9$, corresponding to $\Phi^2=0$. For $\Gamma > 9/2$
the minimum energy for instability is found when $\Psi^2$ is given by
\eq{Psimin}, with the positive sign on the square root. As $\Gamma
\rightarrow \infty$, the minimal value approaches $11/27$, reproducing
the result of Matinyan and M\"{u}ller in the absence of the $\Phi^2/2$
term.

\subsection{Lyapunov Exponents}

The Toda-Brumer test predicts instability above a critical energy, and
other studies show that the coupling between the two fields usually
leads to chaos. However, we wish to confirm this explicitly for the
potential we consider, and to ascertain the parameter values which are
likely to produce cosmologically significant chaos.

We do this by considering the Lyapunov exponents \cite{BakerBK1}.
These are directly related to the fundamental definition of a chaotic
system: sensitive dependence upon initial conditions. A system with an
$n$ dimensional phase space has $n$ Lyapunov exponents. These measure
the logarithm of the expansion of a small $n$-volume of phase space
when it is propagated forwards in time by the equations of motion. In
a chaotic system, this region will be stretched exponentially in at
least one direction, and the signature of this is a positive maximal
Lyapunov exponent. In Hamiltonian systems, the total phase space
volume is conserved, leading to a spectrum of Lyapunov exponents that
has the form $\lambda_1, \lambda_2,\cdots , -\lambda_2,-\lambda_1$,
\cite{OttBK1}.  A corollary is that if a Hamiltonian system is not
chaotic all of its Lyapunov exponents are zero. Since the Lyapunov
exponents come in pairs with the same magnitude and opposite signs we
can check that the values of the exponents we compute numerically are
self-consistent.

We calculate the Lyapunov exponents using the algorithm developed by
Wolf {\em et al.\/} \cite{WolfET1985a} which produces the full
spectrum of $n$ exponents at the expense of solving $n^2+n$
differential equations. In our case, $n=4$, so the cost of calculating
all the exponents is not prohibitive. However, Lyapunov exponents are
defined by a limit as $t\rightarrow\infty$ but the numerical
integrations only run for a finite time. Hence our numerical estimates
of the Lyapunov exponents are never precisely zero in non-chaotic
cases. In practice, we continue the numerical integrations long enough
to ensure that if $\lambda_1>0.005$ and $\lambda_1$ and $|\lambda_4|$
differ by less than 1\% the system is unambiguously chaotic.

\begin{figure}[htbp]
\begin{center}
\psfrag{xlab}[][]{$\Gamma$}
\psfrag{ylab}[][]{$E$}
\psfrag{zlab}[l][]{$\lambda_1$}
\includegraphics[scale=.6]{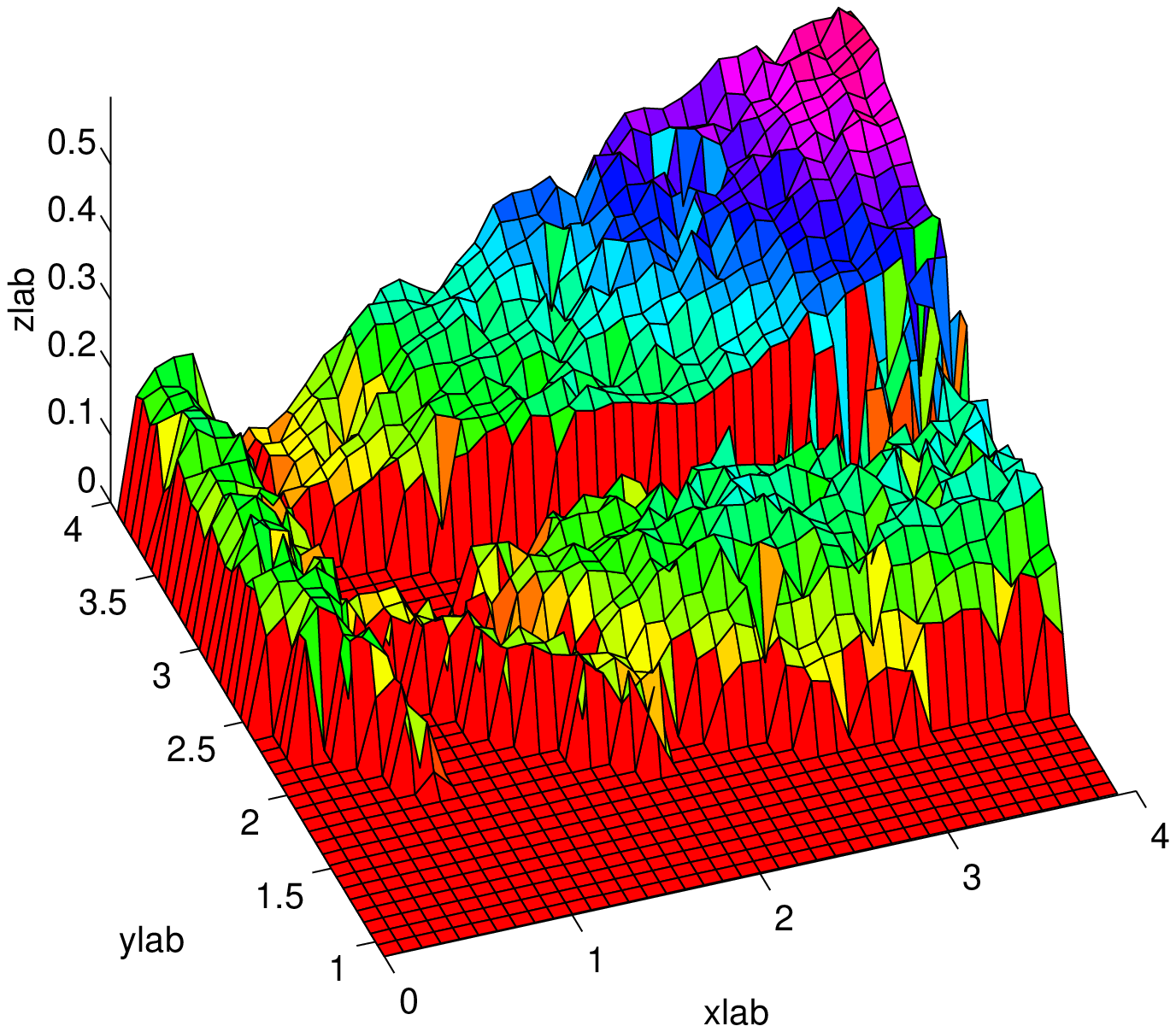}
\end{center}
\caption[]{Maximal Lyapunov exponent, $\lambda_1$ as a function of
energy and $\Gamma$. The specific initial data chosen for the
integrations was $\Phi=1$, $\Psi=1/\sqrt{2}$ and
$\dot{\Psi}=0.64\dot{\Phi}$. The magnitude of the velocities is then
fixed by $E$, and their signs were chosen to be positive.
\label{lyap1}}
\end{figure}

\begin{figure}[htbp]
\begin{center}
\psfrag{xlab}[][]{$\Gamma$}
\psfrag{ylab}[l][]{$E$}
\includegraphics[scale=.7]{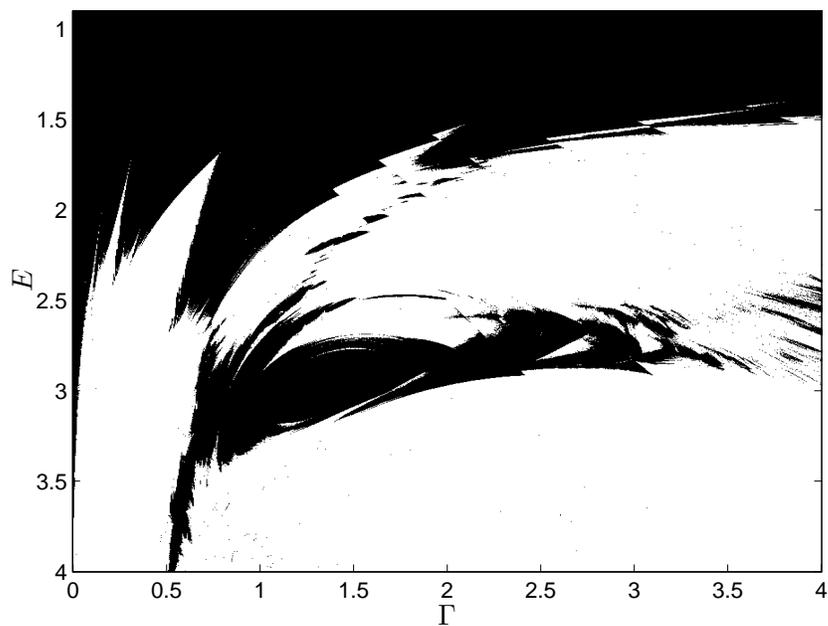}
\end{center}
\caption[]{Region of parameter space for which chaos was detected
(white points), with the same parameters as used in \fig{lyap1}, for
a grid of $840\times840$ equally spaced points.
\label{lyap2}}
\end{figure}

The maximal Lyapunov exponent as a function of $E$ and $\Gamma$ is
displayed in \fig{lyap1}, while \fig{lyap2} shows the region of
parameter space for which chaos is observed.  If $\Gamma\lsim0.05$,
$\Phi$ and $\Psi$ are weakly coupled and the system is typically not
chaotic. However, when $E\gsim2$ and $\Gamma\gsim0.05$ then we usually
(but not always) find chaos. We very rarely found chaos with $E\lsim1$
and never observed chaos with an energy below the minimum value for
instability obtained from the Toda-Brumer test. Furthermore, the
boundary of the region in parameter space which gave rise to chaos is
not smooth, so a small change in parameter values could bring about a
large change in the qualitative nature of the dynamics.  For fixed $E$
and $\Gamma$ only a subset of all the possible orbits in phase space
will exhibit sensitive dependence, so the precise forms of
\figs{lyap1}{lyap2} depend on the chosen initial data.

\section{Chaotic Motion and Inflation}

Having demonstrated the existence of chaos in the Hamiltonian system
derived from the two-field potential, \eq{Vham}, we now turn to the
corresponding inflationary model based on the potential,
\eq{potential}.  Because of the terms proportional to $H$ in
\eqs{phiddot}{psiddot}, the energy density is strictly decreasing. If
the energy changes slowly compared to the typical oscillation time of
the fields, the friction term is a small perturbation to the
Hamiltonian system.  Since there is a lower bound on the energy at
which the potential, \eq{Vham}, can exhibit chaotic motion, any
chaos seen in the inflationary dynamics is transient, switching off
once the energy drops below the critical value.

In general, the oscillatory phase begins with $\psi\sim0$ and
$\phi^2\sim\phiinst^2$, although inflation may cease long before this
condition is fulfilled. We define $\Einst$, the (potential) energy at
the beginning of the oscillatory epoch,
\begin{equation}
\Einst = V(\phiinst,0) = M^4 \left(
  1  + \frac{\sqrt{\lambda}m^2}{\gamma M^2}\right)  =
  M^4 \left( 1 + \frac{1}{\Gamma}\right).
\label{Einst}
\end{equation}
where $\Gamma$ is given by \eq{newvars}. The rate of change in the
energy density is
\begin{equation}
\frac{d E}{dt} = \frac{d}{dt} \left(
 \frac{\dot{\phi}^2}{2}+ \frac{\dot{\psi}^2}{2} + V(\phi,\psi)
 \right) = -3H(\dot{\phi}^2 +\dot{\psi}^2),
\end{equation}
where the second equality uses the equations of motion. During the
oscillatory phase $(\dot{\phi}^2 +\dot{\psi}^2) \sim E$. If we assume
that $M\sim m$, the characteristic oscillation period of the fields
when $E<\Einst$and $E\sim M^4$ will be on the order of $1/M$. Thus
$\Delta E$, the energy lost during a typical oscillation, is roughly
\begin{equation}
\frac{\Delta E}{E} \sim \sqrt{24 \pi} \frac{M}{\mpl}.
\end{equation}
Even if $\Einst$ is only a few times larger than $M^4$, and $M\sim
m\approx 10^{-6}\mpl$, the fields will make tens of thousands of
oscillations as the energy decreases from $\Einst$ to $M^4$, when
symmetry breaking occurs. This estimate is confirmed by direct
numerical computation. Thus the friction term is much smaller than the
other terms in the equations of motion, and the energy loss per cycle
can easily be as low as 1 part in $10^5$. It is likely that the chaos
seen in the Hamiltonian dynamics will survive the addition of such a
small perturbation.

We now turn our attention to the phenomenological aspects of chaos in
the post-inflationary universe. We consider two ways in which the
chaotic motion we have observed in the Hamiltonian dynamics could
influence the observable properties of the universe: by enhancing the
production of topological defects in the early universe, or by
introducing stochastic variations into the relationship between length
scales at the present epoch and the horizon size during inflation.

The smallness of the perturbation caused by the friction term creates
a practical difficulty. Tracing the evolution of the fields through
tens of thousands of oscillations is computationally expensive and,
because the underlying system is chaotic, would require an unrealistic
level of numerical precision. This problem is accentuated when we
consider the evolution for many different choices of initial
conditions. Thus we wish to make the friction more efficient while
still ensuring that it is sub-dominant, so that the numerical
calculations are tractable. We could do this by adding a coupling
between the fields and radiation, in a manner analogous to that
discussed by Albrecht {\em et al.\/} \cite{AlbrechtET1982b}.
Alternatively, we can consider oscillatory motion at an energy scale
considerably higher than that consistent with COBE so that $H$, which
is proportional to the the square root of the energy density, is
larger and the damping more efficient~\cite{AlbrechtET1987a}.  Both
approaches reduce the length of the chaotic era, and therefore reduce
the impact of the chaotic dynamics. We adopt the latter technique, as
it does not alter the structure of the equations, and work with
$M=m=1$, $\lambda=1$, $\gamma=.5$ and $\mpl=5\times 10^{3}m$. We begin
our integrations with $E\approx\Einst=3M^4$. The typical duration of
the oscillatory era is a few hundred oscillation times.  The growth of
the universe during the oscillatory era is not significantly changed
by the scaling.

\subsection{Defect Formation}

Traditionally, only inflationary models which ended with a first order
phase transition were thought to produce topological defects at the
end of inflation. One of the new features of two-field inflation was
that it can easily produce topological defects after a second order
transition, provided the potential has more than one distinct vacuum,
like the case we consider here.

\begin{figure}[htbp]
\begin{center}
\begin{tabular}{c}
\psfrag{xlab}[t][]{$\dot{\phi}(0)$}
\psfrag{ylab}[b][]{$\dot{\psi}(0)$}
\includegraphics[scale=.7]{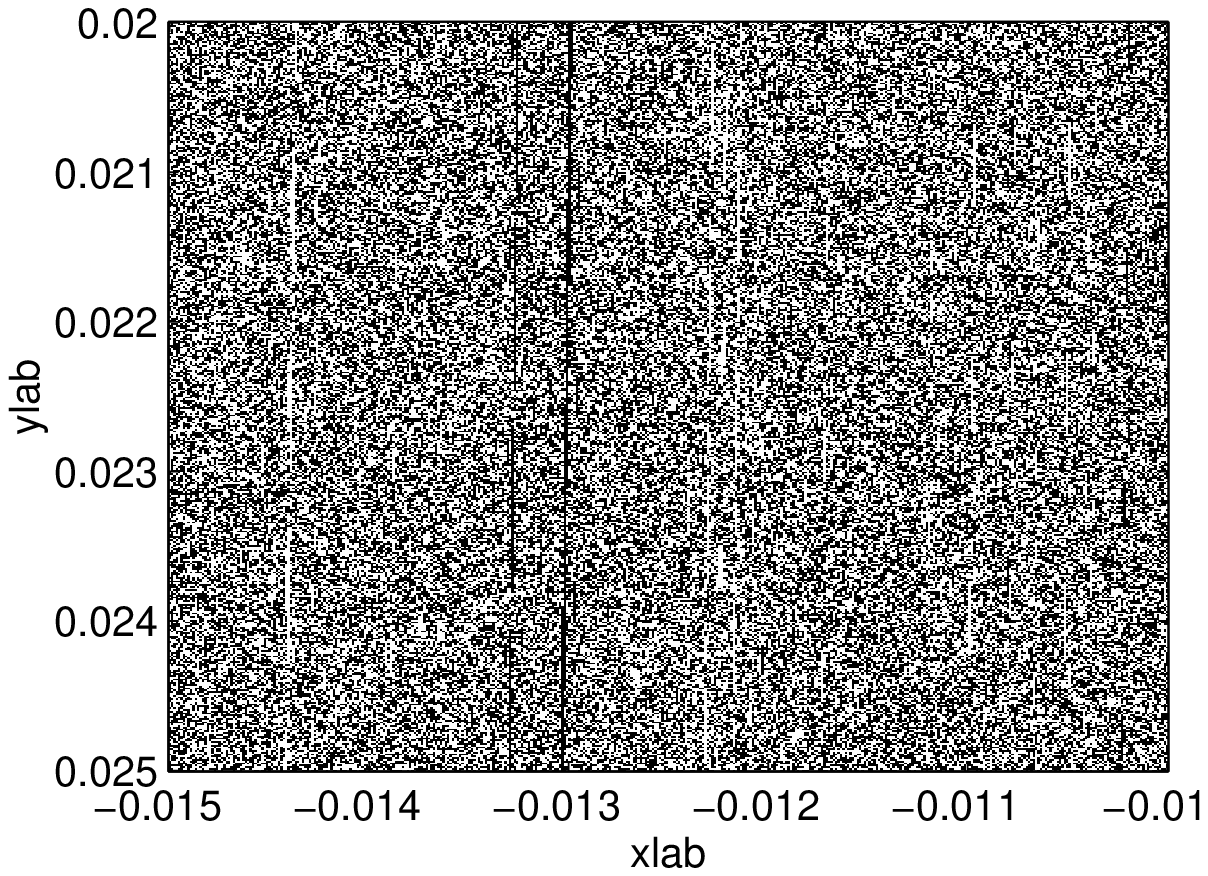}\\ \mbox{} \\
\psfrag{xlab}[t][]{$\dot{\phi}(0)$}
\psfrag{ylab}[b][]{$\dot{\psi}(0)$}
\includegraphics[scale=.7]{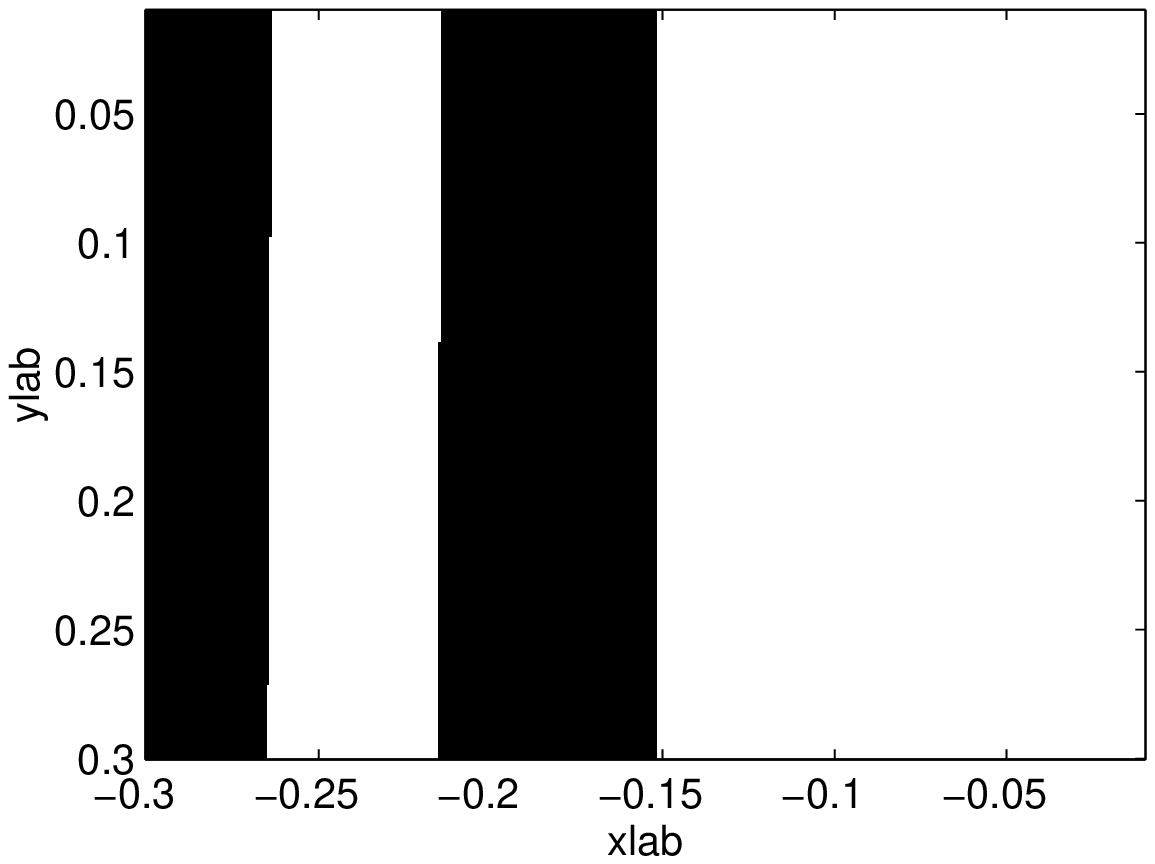}
\end{tabular}
\end{center}
\caption[]{The minima of the potential reached at late times, as a
function of initial conditions (for $420\times420$ equally spaced
points), with $\phi(0)=\phiinst$ and $\psi(0)=0$. White points evolve
towards the vacuum with $\psi>0$. The upper plot shows the results for
$\gamma=0.5$, and those for $\gamma=0$ are shown in the lower
plot. Note the much larger range of initial data considered in the
$\gamma=0$ case. We have chosen $\dot{\phi}(0)$ negative so that
$\phi$ is initially rolling ``downhill''.
\label{defects}}
\end{figure}

In \fig{defects} we show the relationship between the sign of $\psi$
field at late times and the ``initial'' conditions that apply when
$E\approx\Einst$.  We have to chosen to vary the initial velocities,
while keeping the other parameters fixed.  With $\gamma=0.5$ we see
that even very small changes in the initial conditions change the
minimum of the potential picked out by $\psi$ after symmetry breaking.
Conversely, with $\gamma=0$ large changes in the initial data are
required to alter the final state, and there are sharp boundaries
between regions of initial conditions space leading to different
minima.  This qualitative difference is attributable to the fact that
with $\gamma=0.5$ we have transient chaos, whereas in the absence of
friction the underlying dynamics are integrable with $\gamma=0$.

The difference in initial energy between the different configurations
with $\gamma=0.5$ is $\delta E /E \lsim 10^{-4}$, slightly more than
that suggested by COBE, so the range of initial densities is not
unreasonably large.  The spread in energies for the initial conditions
examined with $\gamma =0$ is much wider, giving $\delta E/E
\lsim 5\times 10^{-2}$.

We have chosen to work with parameters that lead to a much shorter
oscillatory period than would be found with realistic values, so we
{\em underestimate\/} the impact of the chaotic period.  Furthermore,
if $\phi$ had rolled freely from the end of inflation its velocity
would be much greater than the range of values we considered. This
causes us to underestimate the initial energy and further reduces the
duration of the chaotic era. On the other hand, we have ignored
couplings between the scalar fields and the rest of the matter sector
and spatial gradients in the fields. Incorporating these effects would
provide additional friction which damps the oscillations of the
$\phi$ and $\psi$ and therefore tends to suppress their chaotic
motion.

Because we do not include gradient terms, the plots in \fig{defects}
represent the symmetry breaking pattern for an ensemble of homogeneous
universes (or horizon volumes), rather than for a single universe in
which the field values vary significantly within a single horizon
volume.  Thus, we do not claim that the results obtained here are
directly applicable to the rate of defect formation in the real
universe.  However, it is clear that two correlated horizon volumes
which have not been in causal contact since the end of inflation are
much less likely to remain correlated after symmetry breaking if the
dynamics of the oscillatory phase permit transient chaos. Thus the
chaotic properties of two-field inflation can be expected to enhance
the level of defect production.

These numerical results apply to a specific set of parameters, but in
general the post-inflationary universe will be chaotic if the energy
density at the beginning of the oscillatory period is high enough to
produce chaos in the corresponding frictionless system.  Thus
transient chaos is a possibility if the parameters combine to give
$\Gamma \gsim 0.05$ and the energy density is greater than than
approximately $2M^4$ at the end of inflation. If inflation ends with
the universe in the inflaton dominated regime, $\dot{\phi}^2$ will be
similar in size to $\Einst$, \eq{Einst}, at the beginning of the
oscillatory era and the energy requirement will almost certainly be
satisfied. However, if inflation lasts into the vacuum dominated
regime the energy density at the end of inflation could be close to
$M^4$, which would not be high enough for chaos even if $\Gamma$, the
rescaled coupling between the fields, was large.

\subsection{The Growth of the Scale Factor}

To make predictions from inflationary models we need to match comoving
distances in the present universe to the comoving horizon size during
inflation. The precise form of this relationship depends on the
thermal history and effective equation of state that applies after the
end of inflation \cite{LiddleET1992b,Turner1993a}.  Derivations of
this relationship typically assume that the many post-inflation
horizon volumes contained within the visible universe have grown by
the same amount since inflation stopped.  This assumption seems
reasonable, as microwave background data suggests that the primordial
universe is homogeneous to within 1 part in $10^{-5}$.  However, a
period of transient chaos can massively amplify small differences in
initial conditions, so we now explicitly examine this assumption by
studying the growth of the scale factor during the interval in which
the energy density evolves from $\Einst$ to $M^4$. We use the same
sets of initial conditions considered in the previous subsection.

\begin{figure}[htbp]
\begin{center}
\begin{tabular}{cc}
\psfrag{xlab}[][]{$\dot{\phi}(0)$}
\psfrag{ylab}[][]{$\dot{\psi}(0)$}
\psfrag{zlab}[][]{$a$}
\includegraphics[scale=.45]{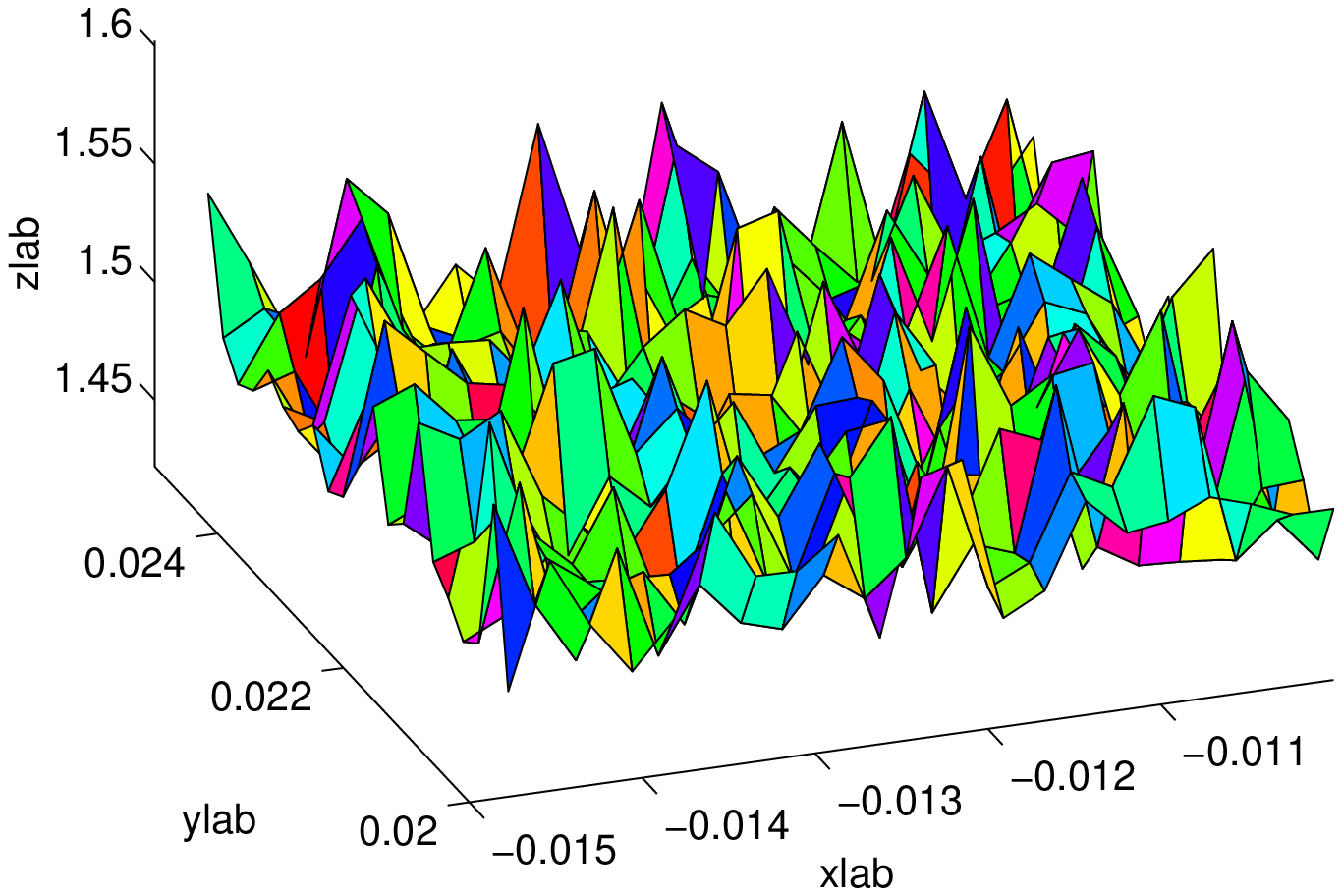}&
\psfrag{xlab}[][]{$\dot{\phi}(0)$}
\psfrag{ylab}[][]{$\dot{\psi}(0)$}
\psfrag{zlab}[][]{$a$}
\includegraphics[scale=.45]{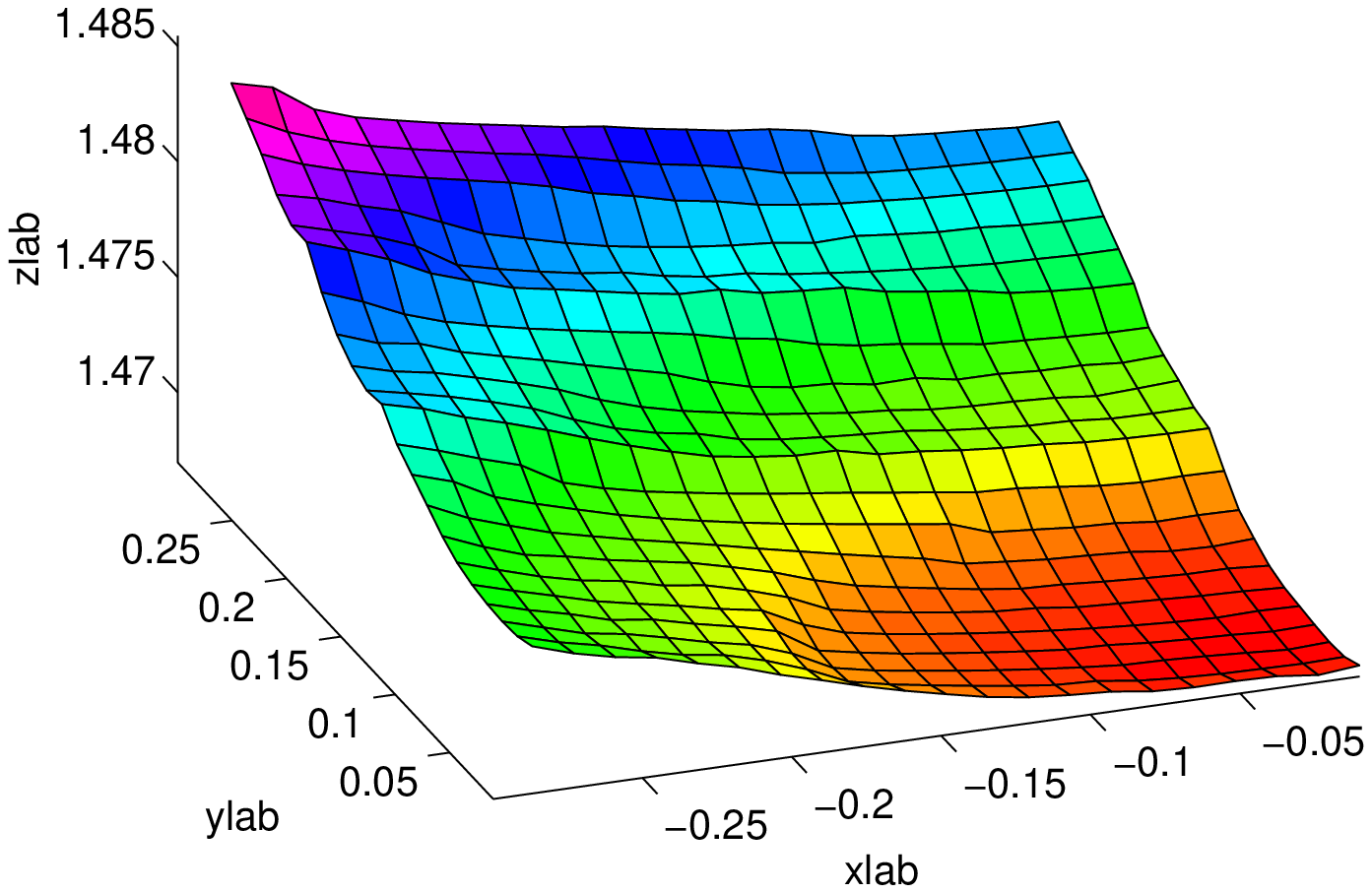}
\end{tabular}
\end{center}
\caption[]{The scale factor at the moment the energy density becomes
equal to $M^4$ and symmetry breaking occurs is plotted as a function
of initial conditions, for $\gamma=0.5$ (left) and $\gamma=0$
(right). The scale factor, $a$, is normalized to be unity at the
beginning of the integration. The initial data are the same as those
used for \fig{defects}.
\label{crossa}}
\end{figure}

In \fig{crossa} we plot the magnitude of the scale factor, $a$, when
$E=M^4$, where $a=1$ when $\phi=\phiinst$.  In the chaotic case, an
initial variance in the energy of less than .01\% leads to differences
of over 10\% in the size of the universe at symmetry breaking.  In the
non-chaotic case the growth of the universe varies much less
dramatically as the initial velocities of the fields are altered, and
what difference we do see is attributable to the variation in initial
energy corresponding to the different initial velocities. Moreover, in
the non-chaotic case the growth of the universe between the start of
the integration and symmetry breaking is a smooth function of the
initial data, while for the chaotic case it is not.

\begin{figure}[htbp]
\begin{center}
\psfrag{xlab}[][]{$t$}
\psfrag{ylab}[][]{$\psi$}
\includegraphics[scale=.8]{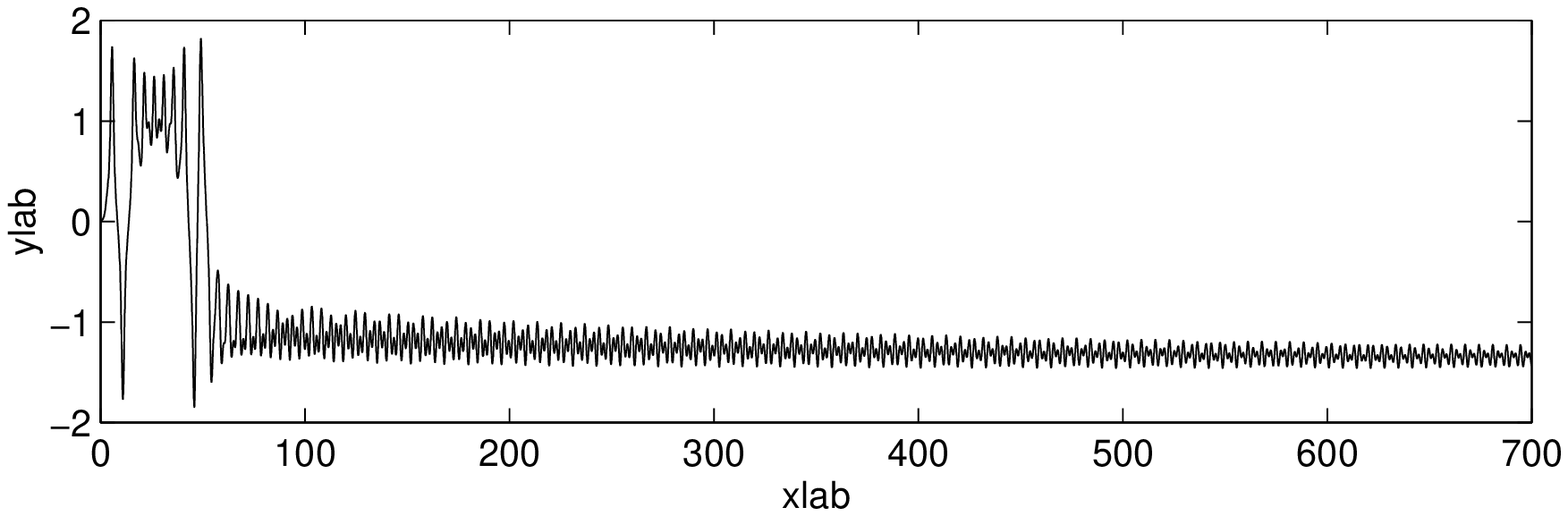}
\psfrag{xlab}[][]{$t$}
\psfrag{ylab}[][]{$\psi$}
\includegraphics[scale=.8]{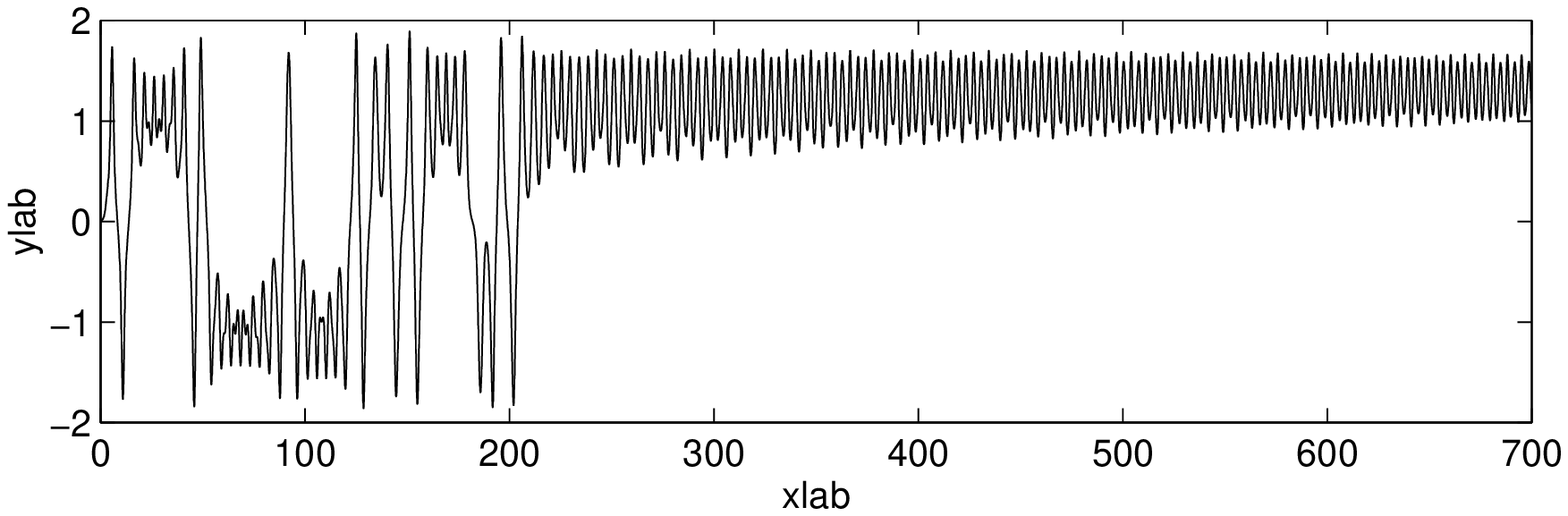}
\end{center}
\caption[]{The evolution of $\psi$ is plotted for two different
initial conditions. In both cases the parameter values are the same as
those used for the chaotic solutions in \fig{crossa}, with the
specific initial initial conditions $\dot{\psi} = .02$ and
$\dot{\phi}=-0.01(1+ 1/419)$ (upper) and $\dot{\phi}=-0.01(1+2/419)$
(lower). The particular numerical values have been chosen to reproduce
specific points in \fig{crossa}. \label{sol}}
\end{figure}

The chaotic case produces large variations in the scale factor because
the precise mixture of kinetic and potential energy varies
significantly with small changes in the initial conditions, as the
fields settle into different meta-stable oscillatory modes.  In turn,
this alters the effective equation of state and the amount of growth
that occurs as the energy density drops from $\Einst$ to $M^4$.  This
is illustrated by \fig{sol}, where the evolution of $\psi$ is plotted
for two slightly different sets of initial conditions.  Obviously, the
solutions diverge after a few oscillations, which is a consequence of
the chaotic dynamics.  Less dramatically, both solutions settle into a
quasi-stable oscillatory mode, but the amplitude of the oscillations
differs significantly between the two solutions.  It is the latter
difference that changes the average kinetic and potential energies.
This effect does not rely on the potential having more than one
vacuum, which is a prerequisite for the formation of topological
defects, but could occur in any two-field model with transient chaos.

In practice, the variation in the scale factor due to chaotic motion
occurring after inflation could introduce a stochastic term into
estimates of the growth of the universe since the end of inflation. In
turn, this would affect estimates of the power associated with the
primordial fluctuations on different length scales, assuming that the
fluctuations generated during inflation differ from the perfectly
scale-free Harrison-Zel'dovich spectrum.  The typical length over
which the scale factor varies will be on the order of the horizon size
at the end of inflation, which is vastly smaller than astrophysical
length scales in the present universe.  However, models which allow a
short burst of inflation after the main period of inflation is over,
such as thermal inflation \cite{LythET1995c} or more complicated
two-field models \cite{RobertsET1994a}, could conceivably magnify the
post-inflationary horizon volumes to the point where the stochastic
properties of the scale factor would not be smeared out on
astrophysical lengthscales.

\section{Discussion}

We have shown that the dynamics of the post-inflationary era in
two-field inflation can be chaotic. We initially analyzed the
evolution of the fields in the frictionless limit corresponding to a
static spacetime background. This reduced the equations of motion for
the fields to a two-dimensional Hamiltonian system. By employing the
Toda-Brumer test and calculating the Lyapunov exponents we
demonstrated that the frictionless system is chaotic.  This approach
is free from many of the difficulties associated with chaos in general
relativity, especially the co-ordinate dependence of the Lyapunov
exponents, as discussed by Rugh~\cite{Rugh1994a}.  We then
demonstrated that the damping term induced by the expansion of the
universe can be viewed as a small perturbation to the
post-inflationary dynamics, and that the most important qualitative
property of chaos - sensitive dependence on initial conditions -
persists when the fields evolve in an expanding universe.  The chaotic
era is transient and lasts from the beginning of the oscillatory
period, which commences when the effective mass squared of the $\psi$
field becomes negative, until the energy density, which is strictly
decreasing, drops below the critical value needed to sustain chaos.

Our analysis is based on a specific potential, which we chose because
its inflationary properties had already received widespread attention,
and because it allowed us to investigate the formation of topological
defects after the end of inflation.  From the perspective of the
idealized, frictionless dynamics chaos is endemic in two-field models
with couplings like $\gamma\phi^2\psi^2$.  Thus we expect that the
type of chaos described here could arise in any inflationary model
with two coupled fields which undergo lightly damped oscillations.

In models of reheating and particle production based on parametric
resonance the inflaton field is coupled to another boson, usually a
scalar, and the resulting system is similar (and sometimes identical)
to the two-field model considered in this paper.
Refs~\cite{KofmanET1997a,GreeneET1997a,GreeneET1997b} show that
particle production during reheating can depend delicately upon the
model parameters, which is the qualitative behavior associated with
chaos. These calculations expand the field(s) in terms of their
spatially dependent modes, so the underlying dynamical system is more
complex than the homogeneous case discussed here. However, the link
between the chaos we have demonstrated in the homogeneous equations of
motion and the phenomenology of reheating is extremely intriguing.

In practice there are several ways in which chaos might be suppressed
in a realistic model of two-field inflation; for instance, the
parameter values could ensure that the oscillatory period does not
begin until the energy density is too low to permit chaos, or the
couplings between the scalar fields and other parts of the matter
sector may ensure that the oscillations are always strongly
damped. Conversely, our numerical integrations were performed with an
artificially large damping term, and the parameter values we chose
ensured that the oscillatory period did not begin until the energy
density was close to the height of the barrier between the two vacua
in the potential. Restoring a realistic damping term or increasing the
energy density at which oscillations begin would make the chaotic
period last much longer than it does in the case considered here.

We have not included spatial variations in the fields, or couplings
between the scalar fields and other forms of matter.  Thus we cannot
determine whether a chaotic period necessarily leads to detectable
effects, and if these effects are useful to the model builder, or in
conflict with observational evidence. However, our work shows that it
is feasible for chaotic effects to influence the observable properties
of two-field models. Thus, approximations which do not preserve the
chaotic properties of the full system could lead to spurious
conclusions about the phenomenology of two-field inflation.  It will
obviously be important to investigate these issues in more detail.

Inflationary models based on a single scalar field in a spatially flat
background do not possess enough degrees of freedom to become
chaotic. Since spatial curvature is negligible after inflation, a
chaotic era at the end of inflation requires two or more scalar
fields. Thus any distinctive observable effects associated with the
period of transient chaos we have identified could provide a mechanism
for testing a broad class of two-field models, in the same way that
constraints on the scalar and tensor perturbation spectra test a large
group of models with a single slowly rolling field
\cite{LidseyET1995a}.  Furthermore, chaotic effects are often
associated with fractal structures. Due to the complex nonlinear
dynamics associated with defect formation, cosmic string networks have
a fractal geometry \cite{VachaspatiET1984a,Pagels1987a}, while
stochastic inflationary inflationary models have a fractal structure
at scales much larger than a typical horizon \cite{AryalET1987a}.
Thus we speculate that a chaotic era at the end of inflation may give
rise to fractal structure, in which case two-field models of inflation
could make testable predictions about the fractal properties of the
universe at large scales.

We believe that the scenario analyzed here is the most interesting
example of chaotic dynamics in an inflationary model to date.  We have
found chaos in a model that has already been widely studied because of
its promising phenomenology, and we did not need to include
non-minimal couplings or non-zero spatial curvature to ensure the
presence of chaos.  Because of the sensitive dependence of chaotic
systems, small differences in the initial conditions which apply at
the beginning of the oscillatory epoch can have a significant impact
on the subsequent evolution, affecting ``zero order'' quantities such
as the scale factor and the minimum of the potential that the fields
evolve towards.  Moreover, because the chaotic era occurs after the
end of inflation its effects might, in principle, be detectable by
human observers.

\section*{Acknowledgments}
This work was partially supported by a Grant-in-Aid for Scientific
Research Fund of the Ministry of Education, Science and Culture
(Specially Promoted Research No. 08102010), and the Waseda University
Grant for Special Research Projects. RE is supported by DOE contract
DE-FG0291ER40688 (Task A), and thanks Cornell University for its
hospitality while part of this work was carried out. Computational
work in support of this research was performed at the Theoretical
Physics Computing Facility at Brown University. We thank Robert
Brandenberger and David Wands for their comments on a draft of this
paper.

\newpage

\end{document}